\documentclass[12pt]{article}

\usepackage{amsfonts}
\usepackage{amsmath}

\usepackage{amssymb}

\usepackage[vcentermath,enableskew]{youngtab}

\newcommand{\I}{\imath}

\newcommand{\ID}{\mathbf{1}}

\newcommand{\CAS}{\mathrm{C}_2}
\newcommand{\V}{\mathcal{V}}

\newcommand{\CP}{\mathbb{C}\mathrm{P}}

\begin{document}

\title{ Scalar Field Theory on Fuzzy $S^4$            }
\author{ 
Julieta Medina\footnote{Email: julieta@stp.dias.ie }
\; {\it and}\; 
Denjoe O'Connor\footnote{Email: denjoe@stp.dias.ie}
\\ 
\\
{\it School of Theoretical Physics},\\
\it{ Dublin Institute for Advanced Studies},\\ 
{\it 10 Burlington Road, Dublin 4}, 
{\it Ireland} \\
\\   and \\ 
\\
{\it Depto de F{\rm\'\i}sica, Cinvestav, Apartado Postal 70-543,} \\
{\it {M\'exico D.F. 0730, M\'exico}}\\ \hfill\\
\\ }

\maketitle

\begin{abstract}
Scalar fields are studied on fuzzy $S^4$ and a solution is found
for the elimination of the unwanted degrees of freedom that occur in the
model. The resulting theory can be interpreted as a Kaluza-Klein reduction
of $\CP^3$ to $S^4$ in the fuzzy context.
\end{abstract}
\vfill\eject
\section{Introduction}

To access the physics of field theories in the strong coupling regime
non-perturbative methods are necessary. These typically involve the
reduction of the field theory to a model with a finite number of
degrees of freedom and the use of numerical methods to sample the
possible configurations. The now standard method in this regard is
lattice field theory.  It has been developed in the last twenty years
into a refined tool \cite{Montvay_Munster_book}.

However, resorting to a lattice is not the only possible method of
reducing a field theory to a finite number of degrees of freedom. An
alternative is what has become known as the fuzzy approach
\cite{Madore}-\cite{kimura}, see \cite{Pramana} for a review.


In this approach one takes the underlying space in the form of its
algebra of functions and seeks a sequence of non-commutative algebras
with finite dimensional representations, whose limiting form
reproduces the commutative algebra. The elements of the algebra can
then be represented by matrices and will play the role of scalar
fields in an approximation to field theory.  Dirac and Yang-Mills
fields on fuzzy spaces have also been considered.  These involve
projective modules over the algebra \cite{chiral_anomaly}. The method
has further been extended to superspace and a fuzzy supersphere has
been constructed \cite{supersphere}.

As a non-perturbative approach to field theory the fuzzy scheme is
very different from the lattice one and many new features emerge.  The
most surprising is perhaps the phenomenon of UV/IR mixing
\cite{Vaidya,Brian_Denjoe_Peter,Chu_etal} where a residue of the
microscopic non-commutativity remains in the commutative limit. This
can be eliminated by suitably modifying the original action of the
model.

The most intensely studied example in the fuzzy approach is the so
called fuzzy sphere, $S^2_F$, which is realized as the matrix algebra
of dimension $L+1$, denoted here $Mat_{L+1}$, with the inner
product\footnote{ Note that $(M,M)=||M^\dag M||=||M||^2$ where
$||\cdot||$ is the $C^*$ norm on $Mat_{L+1}$.}
$(M,N)=\frac{Tr}{L+1}(M^\dag N)$, where $M$ and $N \in Mat_{N+1}$ and
the geometry is specified through the set of derivations ${\cal
L}_i=Ad L_i$ that correspond to the adjoint action of the generators
$L_i$ of SU(2) (with spin $s=L/2$).  For scalar fields only the
Laplacian (or sequence of Laplacians parametrized by the matrix size)
plays a role and the geometry is specified by a choice of Laplacian.
Choosing the Laplacian to be ${\cal L}^2$ gives us a round sphere.
One can therefore view the round fuzzy sphere, $S^2_F$, in the spirit
of Connes \cite{Connes_book} as the triple ($Mat_{L+1}$, ${\cal L}^2$, $(\cdot,\cdot)$)
i.e. given by the algebra $Mat_{L+1}$ together with the differential
operator ${\cal L}^2$ and the inner product
$(M,N)=\frac{Tr}{L+1}(M^\dag N)$.  More generally one could consider a
non-round sphere by utilizing an alternative expression for the
Laplacian.

For $S^2_F$ the algebra $Mat_{L+1}$ can be viewed as generated by
$X_i=\frac{R}{\sqrt{s(s+1)}}L_i$, which satisfy
\begin{equation}
\sum_{i=1}^3X_i=R^2\ID \quad .\label{s2_F_defn}
\end{equation}
The $X_i$ satisfy the commutation relations
\begin{equation}
[X_i,X_j]=\frac{iR}{\sqrt{s(s+1)}}\epsilon_{ijk}X_k
\end{equation}
and become commutative in the large $L$, fixed $R$, limit so that we
recover the commutative algebra of functions on $S^2$. If instead one
scales $R$ with $L$ one can obtain the non-commutative plane
\cite{Chu_etal,Sachin_Badis}.

The construction above can be carried out for any $\CP^N$ \cite{starprod_CPN}
and will be reviewed from the above point of view in the next section.
One can imagine carrying the prescription out more generally by
specifying the family of matrices and Laplacians such that one
recovers the approximated space up to isospectral
equivalence. However, examples beyond the most symmetric spaces have
not yet been constructed in this way.

The known constructions (to date) of matrix approximations to
continuum spaces are almost exclusively coadjoint orbits and their
products. The one apparent exception\footnote{See also the related
construction of fuzzy even \cite{even} and odd \cite{odd} spheres.} is
$S^4$ (see
\cite{grosse_peter,castelino,Ramgoolam_0105006,kimura}). This is an
especially important example since it is the most natural replacement
of ${\mathbb R}^4$ in studies of Euclidean quantum field theory.  It
is also unusual in that $S^4$ does not admit a symplectic structure
and hence a fuzzy version of it could not be achieved by quantization
of the classical space. It is, therefore, important to clarify in what
sense a matrix approximation to $S^4$ exists.

We will demonstrate that the matrix algebra approximation of $S^4$
under discussion is really a matrix version of $\CP^3$ in
disguise. The desired $S^4$ emerges from a Kaluza-Klein type
construction in the fuzzy context.

Curiously, the space $\CP^3$ has recently been considered \cite{Bernevig_etal}
in the context of higher dimensional quantum hall liquids and perhaps some
of the techniques developed here may be of use in this context. 

The structure of the paper is as follows: In section 2 we present an
overview of the fuzzy approach in the case of $\CP^N$. In section 3 we
show that $S^4$ emerges from a matrix construction as the matrix size
is sent to infinity.  In section 4 we study the representation content
of the matrix algebra and give our solution to the problem of
suppressing non-$S^4$ modes. As a side product of our construction we
are able to find the projector necessary to eliminate non-$S^4$
modes. Section 5 gives our prescription for the scalar field action
where the non-$S^4$ modes are dynamically suppressed. One can see from
the general discussion that it corresponds to a $\CP^3$ model where
the maximal $SO(6)$ symmetry is reduced to an $SO(5)$
symmetry. Geometrically the model corresponds to a Kaluza-Klein type
space which is an $S^2$ bundle over $S^4$ with the radii of the $S^2$
fibres being sent to zero as the energy scale of the unwanted modes is
sent to infinity.  Section 6 gives our conclusions.

\section{Construction of $\CP^N_F$}

We begin by reviewing the construction of $\CP^{N}_F$. For this one
takes the $L$ fold symmetric tensor product of the fundamental
representation of $SU(N+1)$, which in terms of Young tableaux is the
denoted
\begin{equation}
\mathbf{d^N_{L}} = 
\Yvcentermath1{{\tiny{\underbrace{
  \yng(1)\dots \yng(1) }_{L}}}}	
\end{equation}
where $d^N_{L}=\frac{(L+N)!}{N!L!}$.  The sequence of matrix algebras
under consideration will then be $Mat_{d^N_L}$ which will be endowed
with the inner product $(M,N)=\frac{Tr}{d^N_L}(M^\dag N)$ with $M$ and
$N \in Mat_{d^N_L}$.

The geometry has not as yet been specified since a sequence of
differential operators is not yet given. When the geometry is
specified by a Laplacian, the number of eigenvalues of that Laplacian,
less than a cutoff momentum scale $\Lambda$, increases with the cutoff
as $\Lambda^d$. Since the total number of degrees of freedom in the
matrix algebra is $(d^N_L)^2$ and our cutoff is $L$, we can deduce the
dimension of the space being approximated via
\begin{equation}
d=2\lim_{L\rightarrow\infty} \frac{\ln d^N_L}{\ln L}=2N
\label{dim_cpn}
\end{equation}
which is consistent with $\CP^N$. 

If we choose our differential operators to be built from the
generators, $T_a$, of $SU(N+1)$ in the $\mathbf d^N_L$ representation
and the generators $-T_a^R$ in the complex conjugate representation,
with these latter operators acting on matrices on the right, then
matrices are equivalent to the product representation
\begin{equation}
{\tiny\Yvcentermath1
  \underbrace{\yng(1)\dots \yng(1)}_{L}\otimes 
\overline{\underbrace{\yng(1)\dots\yng(1)}_{L}}}
\label{young_matrices}
\end{equation}
which can be expanded in terms of $SU(N+1)$ representations to yield
the expansion of matrices in terms of polarization tensors.
For example for $SU(4)$ with $L=3$ we have 
$$\overline{{\tiny\Yvcentermath1  \yng(3)}}={\tiny \yng(3,3,3)}$$
and the expansion:
\begin{equation}
{\tiny\Yvcentermath1
  \yng(1,1,1)\otimes \yng(3,3,3)
=
\ID \oplus \tiny{\yng(2,1,1) \oplus \yng(4,2,2) \oplus \yng(6,3,3)}}
\label{young_su_4}
\end{equation}
In the special case of $SU(2)$, for example with $L=3$, we have
$${\tiny{\overline{\Yvcentermath1  \yng(3)}= \yng(3)}}$$
and the decomposition
\begin{equation}
\Yvcentermath1 {\tiny{\yng(3)\otimes \yng(3)}}
=
\ID \oplus {\tiny{\yng(2)}} \oplus {\tiny{\yng(4)}} \oplus {\tiny{\yng(6)}}
\label{young_su_2}
\end{equation}
and we see that the expansion is in terms of integer angular momentum
and is cut off at $l=3$. In general for the $L$-fold symmetric tensor
product representation, the angular momentum will be cutoff at $l=L$
and as $L\rightarrow\infty$ we recover all of the representations
corresponding to functions on $S^2$. Similarly in the $L\rightarrow\infty$ 
limit of the $SU(N+1)$ case we recover $\CP^N$.  

Fuzzy $\CP^N$ would then, by analogy with $S^2_F$, be considered as
the triple ($Mat_{d^N_{L}}$, ${\cal L}^2$, $(\cdot,\cdot)$) where now
the sequence of matrices is that of dimension\footnote{To avoid
confusion in the case of $\CP^3$ we sill simply denote $d^3_L$ by
$d_L$.}  $d^N_{L}=\frac{(L+N)!}{N!L!}$, ${\cal L}^2=(Ad T_a)^2$ with as
before $(M,N)=\frac{Tr}{d^N_{L}}(M^\dag N)$, where $M$ and $N\in
Mat_{d^N_{L}}$.

A scalar field action on $\CP^N$ would then correspond to 
\begin{equation}
S[\Phi]=\frac{Tr}{d^N_{L}}[\frac{1}{2}\Phi{\cal L}^2\Phi +V(\Phi)]
\end{equation}
The $\phi^4$ model of this type on $S^2_F$ has been analyzed in
\cite{Brian_Denjoe_Peter} and is currently being studied numerically
as a testing ground for the feasibility of the numerical approach.

\section{Construction of $S^4_F$}
Let us now turn to the construction of $S^4_F$. For this observe that
$X_a=\frac{R}{\sqrt{5}}\Gamma_a$, with $\Gamma_a$ the Dirac matrices
including $\gamma_5$, satisfy
\begin{equation}
\sum_{a=1}^5 X_aX_a=R^2\ID. \label{s4_F_defn}
\end{equation}
This gives us a matrix approximation to the defining equation of $S^4$
in ${\mathbb R}^5$ and is the direct analogue of (\ref{s2_F_defn}) for
$S^2_F$.  To get our sequence of matrices approximating $S^4$ we are
led to consider representations of the group $Spin(5)$ (or
equivalently $Sp(2)$).  In the above we have used the defining four
dimensional representation $(\frac{1}{2},\frac{1}{2})$ of
$Spin(5)$. We can therefore consider the irreducible representation
obtained from the $L$ fold symmetric tensor product of this
representation i.e. the $Spin(5)$ representation
$(\frac{L}{2},\frac{L}{2})$ which will contain a set of five matrices:
$J_a$, $a=1,...,5$ which can be realized as the symmetrization of
$L$ copies of the $\Gamma$ matrices in the $Spin(5)$ fundamental
representation:
\begin{equation}
    J_a  =    \left( \underbrace{ \Gamma_a {\otimes} \mathbf{1} 
                            {\otimes} \cdots {\otimes} 
                            \mathbf{1}}_{L-times}   +
                            \mathbf{1} {\otimes}  \Gamma_a
                            {\otimes} \cdots {\otimes} 
                            \mathbf{1}   + \cdots +
                            \mathbf{1} {\otimes} \mathbf{1} 
                            {\otimes} \cdots {\otimes} 
                             \Gamma_a
                  \right)_{sym}.  \label{defJa}
\end{equation}
where the subscript $sym$ indicates that we are 
projecting onto the irreducible totally symmetrized representation.
These matrices satisfy the relation
$$J_aJ_a=L(L+4)\ID$$ so that we can define a sequence of
matrices\footnote{We do not cumber the notation by making the
dependence on $L$ explicit.}, $X_a$, given by
\begin{equation}
X_a=\frac{R}{\sqrt{L(L+4)}}J_a
\label{in.2}
\end{equation}
The algebra generated by these functions will become commutative in
the infinite $L$ limit (see the discussion associated with
eq. (\ref{commutator_of_s4})) and since in the commutative case it is
easy to check that the five co-ordinate functions $x_a$ (satisfying
$x_a^2=R^2$) generate $C^{\infty}(S^4)$ we will recover $S^4$ in the
limit. 

If we return to the matrices arising at level $L=1$, we see that the
$\Gamma_a$ are not sufficient to form a basis for all matrices, but
rather, to get a basis, we need to include the matrices
$\sigma_{ab}=\frac{1}{2i}[\Gamma_a,\Gamma_b]$.  We can now expand any
$4\times4$ matrix in terms of the 16 matrices
$\{\ID,\Gamma_a,\sigma_{ab}\}$.  In this approximation (i.e. $L=1$)
a matrix representing a function on $S^4$ 
will be of the form 
\begin{equation}
F=F_0\ID+F_a\Gamma_a
\end{equation}
and our cutoff angular momentum on $S^4$ will again be $l=1$ (as it
was at the corresponding $L=1$ approximation to $S^2$).  However, a
matrix product of two such functions will involve a non-zero
coefficient of $\sigma_{ab}$ and in the absence of arbitrary such
coefficients the algebra does not close. These parameters will have no
corresponding counterparts in the expansion of functions on
commutative $S^4$. One option as argued by Ramgoolam
\cite{Ramgoolam_0105006} is to project out such terms, in which case
one is left with a non-associative algebra. This involves additional
complications and does not seem particularly suited to numerical
work. In addition the necessary projector must be constructed. We will
return to this point in a concluding section where we will, in fact,
give the projector.

An alternative is to include arbitrary coefficients of $\sigma_{ab}$
(demanding an associative algebra) and attempt to suppress such
coefficients of unwanted terms, by making their excitation improbable
in the dynamics. In this approach our algebra will be a full matrix
algebra and obviously associative.  The principal task of this paper
will therefore be to give a prescription for suppressing the
additional modes that arise in this extended algebra.

From a physics point of view the matrices are to play the role of our 
scalar fields which will be sampled in a Monte-Carlo simulation. 
We will therefore be seeking an appropriate scalar field action
which suppresses the non-$S^4$ modes in a probabilistic sense 
in our simulations.  

A successful method of suppressing the unwanted modes would be to add
to the scalar action a term $S_I[\Phi]$ which is positive for any
$\Phi$ and zero only for matrices that correspond to functions on
$S^4$, and non-zero for those that do not. The modified action would
therefore be of the form $S[\Phi]+h S_I[\Phi]$. The parameter $h$
should then be chosen large and positive.  The probability of any
given matrix configuration then takes the form
\begin{equation}
{\mathcal P}[\Phi]=\frac{{\rm e}^{-S[\Phi]-hS_I[\Phi]}}{Z}
\label{prob_of_config}
\end{equation}
where 
\begin{equation}
Z=\int d[\Phi] {\rm e}^{-S[\Phi]-hS_I[\Phi]}
\label{partition_fn}
\end{equation}
is the partition function of the model.

We will show that this can be achieved by choosing $S_I[\phi]=\frac{Tr}
{d_L}(\frac{1}{2}\Phi\Delta_I\Phi)$ with
$d_L=\frac{(L+1)(L+2)(L+3)}{6}$ and $\Delta_I$ a positive operator. It
can be interpreted as a modification of the Laplacian which is zero on
matrices corresponding to $S^4$.  

From (\ref{dim_cpn}) we see that the dimension of the space being
approximated by this sequence of matrices is in fact six and not
four. We will further see that the entire model, when the unwanted
degrees of freedom are included, corresponds to a fuzzy version of
$\CP^3$.  $\CP^3$ is an $S^2$ bundle over $S^4$ and the parameter $h$
can be related to the radius of the $S^2$ fibres over $S^4$ with the
radius being sent to zero as $h\rightarrow\infty$
\cite{Brian_Julieta_Denjoe}.  In this sense we will have a fuzzy
Kaluza-Klein type space whose low energy limit is $S^4$.

\section{The representation content and Laplacian}

The matrices $\frac{\sigma_{ab}}{2}$ with $a,b=1,\dots,5$ are the
generators of $Spin(5)$ in the fundamental representation. If we
further identify $\Gamma_a=\sigma_{a6}=-\sigma_{6a}$, then the set
$\frac{\sigma_{AB}}{2}$ with $A,B=1,..,6$ are the generators of
$Spin(6)$ in its fundamental representation
$(\frac{1}{2},\frac{1}{2},\frac{1}{2})$. We can similarly identify
$J_{AB}$ in the $L$ fold symmetric tensor product representation, by
replacing $\Gamma_a$ in (\ref{defJa}) by $\frac{\sigma_{AB}}{2}$. The
resulting set $J_{AB}$ satisfy the algebra:
\begin{equation}
  \left[J_{AB}, J_{CD}\right] = \I \left(\delta_{AC}J_{BD}+
\delta_{BD}J_{AC}- \delta_{AD}J_{BC}-\delta_{BC}J_{AD} \right)
\label{algSO6}
\end{equation}
and generate the $Spin(6)$ irreducible representation
$(\frac{L}{2},\frac{L}{2},\frac{L}{2})$ with dimension
$d_L=\frac{(L+1)(L+2)(L+3)}{6}$. The subset $J_{ab}$, are $Spin(5)$
generators in the $(\frac{L}{2},\frac{L}{2})$ representation and the
subset $J_{a6}=\frac{J_a}{2}$ transform as a vector under $Spin(5)$ in
this representation.

We can therefore view the $4\times4$ matrix algebra as the tensor
product
$(\frac{1}{2},\frac{1}{2})\otimes\overline{(\frac{1}{2},\frac{1}{2})}$
if we take a $Spin(5)$ point of view or as
$(\frac{1}{2},\frac{1}{2},\frac{1}{2})\otimes
\overline{(\frac{1}{2},\frac{1}{2},\frac{1}{2})}$ from a $Spin(6)$
perspective. Similarly, for the $L$ dependent sequence of
matrices\footnote{ Note, for $Spin(5)$ we have
$\overline{(\frac{L}{2},\frac{L}{2})}= (\frac{L}{2},\frac{L}{2})$
while for $Spin(6)$ we have
$\overline{(\frac{L}{2},\frac{L}{2},\frac{L}{2})}=
(\frac{L}{2},\frac{L}{2},-\frac{L}{2})$.} one can take either a
$Spin(5)$ or $Spin(6)$ view of the matrix algebra. The dimension of
both sequences of representations is $d_L=\frac{(L+1)(L+2)(L+3)}{6}$
and the sequence of matrix algebras under consideration is
$Mat_{d_L}$. From (\ref{dim_cpn}) we see that this sequence is one
associated with an approximation to a six rather than four dimensional
space.  Since $Spin(6)= SU(4)$ the natural geometry associated with
the $Spin(6)$ approach is that of a fuzzy approximation to $\CP^3$.

In fact all the representations under consideration here can also be
considered as representations of $Spin(6)=SU(4)$. Note, the $L$ fold
symmetric tensor product representation
$(\frac{L}{2},\frac{L}{2},\frac{L}{2})$ of $Spin(6)$ is precisely the
representation
\begin{equation}
{\tiny\Yvcentermath1
  \underbrace{\yng(1)\dots \yng(1)}_{L}}
\nonumber
\end{equation}
or equivalently the $(L,0,0)$ representation of $SU(4)$.

In this sequence of representations the $Spin(5)$ generators $J_{ab}$
are still of the form
\begin{equation}
  J_{ab} =\frac{1}{ 4 i} [J_a, J_b ] \quad \mbox{a,b=1,...,5} \label{defJab}
\end{equation}
as can be seen from (\ref{algSO6}). It is easy to verify that
\begin{equation}
J_{ab}J_{ab}=L(L+4)\ID \quad .
\end{equation}
So the commutator of the coordinate matrices defined in (\ref{in.2})
is given by
\begin{equation}
 [X_a,X_b]= 4 i R^2 \frac{J_{ab}}{L(L+4)}\quad . 
\label{commutator_of_s4}
\end{equation}
In the $L\rightarrow \infty$ with $R$ fixed the right hand side of
(\ref{commutator_of_s4}) goes to zero, and the coordinates commutes.
We still have retained the constraint (\ref{s4_F_defn}), we recover 
{\em commutative} $S^4$.

We have not as yet defined a geometry. For this, as discussed above,
we need a Laplacian. From the above discussion is is clear there are
now two available candidates, the quadratic Casimir operator of
$SO(6)$ or that of $SO(5)$, which are respectively:
\begin{eqnarray}
 \CAS^{SO(6)} &=& \frac{1}{2}(Ad J_{AB})^2 \label{so6_cas}\\
 \CAS^{SO(5)} &=& \frac{1}{2}(Ad J_{ab})^2 \label{so5_cas}
\end{eqnarray}
If we chose the $SO(6)$ Casimir (which is equally the $SU(4)$ Casimir)
we are imposing the geometry of a round $\CP^3$. But of course for a
round $S^4$ $SO(5)$ symmetry is all we need.

If we take the $Spin(5)$ point of view then an arbitrary matrix can be
considered as an element of the vector space
$(\frac{L}{2},\frac{L}{2})\otimes
\overline{(\frac{L}{2},\frac{L}{2})}$ which reduces under $Spin(5)$
as:
\begin{equation}
(\frac{L}{2},\frac{L}{2})\otimes
(\frac{L}{2},\frac{L}{2})=
\sum_{n=0}^{L}\sum_{m=0}^{n} (n,m)\quad .
\label{rep_content}
\end{equation}
with the dimension of $(n,m)$ being
\cite{perelomov1} \\
\begin{equation}
dim(n,m)=\frac{1}{6}\left(2n+3\right)\left(2m+1\right)
                      \left(n(n+2) -m(m+1)\right)  \quad .
\end{equation}
Only the representations $(n,0)$ correspond to functions on $S^4$, all
others are non-$S^4$ representations.

Equivalently taking the $Spin(6)$ point of view
one has the $Spin(6)$ reduction
\begin{equation}
(\frac{L}{2},\frac{L}{2},\frac{L}{2})\otimes
(\frac{L}{2},\frac{L}{2},-\frac{L}{2})=
\sum_{n=0}^{L} (n,n,0)
\end{equation}
where
\begin{equation}
dim(n,n,0)=\frac{1}{6}\left(2n+3\right)\left(n+1\right)^2
                      \left(n+2\right)^2 .
\end{equation}
Furthermore, one can see that the $SO(6)$ representation $(n,n,0)$
breaks up as a sum of $SO(5)$ representations and we have:
\begin{equation}
(n,n,0)=\sum_{m=0}^{n}(n,m) \quad .
\end{equation}

One can gain more insight into the role of these representations and
the above decompositions by thinking of the above arguments in terms
of polarization tensors.

In our context, an arbitrary matrix, which plays the role of 
a scalar field, can be decomposed into a sum of orthonormal 
polarization tensors where the set of polarization tensors carry the 
representation content (\ref{rep_content}). Thus an arbitrary matrix
$M \in Mat_{d_L}$, can be decomposed as 
\begin{equation}
 M = \sum_{n=0}^{L} \sum_{m\leq n}  
                M^{(n,m)}_{a_1,a_2,\cdots,a_{n+m}} 
               \V^{(n,m)}_{a_1,a_2,\cdots,a_{n+m}}                  
 \label{Mhat}
\end{equation}
where
\begin{equation}
     \V^{(n,m)}_{a_1,a_2\cdots, a_{m+n}}  \in \hbox{the $SO(5)$ IRR 
$(n,m)  \quad  m \leq n$}
\end{equation}
are the polarization tensors.  These are in fact appropriately
symmetrized $n$-th order polynomial of $J_a$ and $J_{ab}$, with traces
removed and with the order of $J_{ab}$ being $m$. Those that
correspond to functions on $S^4$ are therefore the $\V^{(n,0)}$ and
are matrix versions of the $S^4$ spherical harmonics and the direct
analogues of the $\hat Y_{lm}$ of \cite{Brian_Denjoe_Peter} but in an
orthogonal basis.

In fact, the decomposition (\ref{Mhat}) corresponds to 
the $SU(4)$ decomposition:
\begin{eqnarray*}
    \mathbf{d_L} \otimes \mathbf{\overline{d_L}} & = &
    \mathbf{1}+\mathbf{15} + \mathbf{84} + \cdots +\mathbf{D_{n-1}} +
    \mathbf{D_n} \\
\end{eqnarray*}
where ${\rm D_n}=dim(n,0,n)=\frac{1}{12}(2n+3)(n+1)^2(n+2)^2$
with $\mathbf{D_n}$ further decomposed under $Spin(5)$ as
\begin{equation}
\mathbf{D_n}=\sum_{m=0}^{n}(n,m) \quad .
\end{equation}
For example the $\mathbf{15}$ breaks up as
$\mathbf{15}=\mathbf{5}+\mathbf{10}$ which corresponds to the
decomposition of $J_{AB}$ into $J_a$ and $J_{ab}$.  Similarly the
$\mathbf{84}$ decomposes as
$\mathbf{84}=\mathbf{14}+\mathbf{35}+\mathbf{35}'$ and so on.

Let us now discuss the eigenvalues of the Casimirs on the above
representations. We have
\begin{eqnarray}
\CAS^{SO(6)} \V^{(n,m)}             
         & = & 2n(n+3) \V^{(n,m)}  \label{D100} \\
\CAS^{SO(5)}\V^{(n,m)}  
         & = &\left\{ n(n+3)+m(m+1)\right\}\V^{(n,m)}  \quad .
\label{D101} 
\end{eqnarray}
We see that the operator
\begin{equation}
\mathrm{C}_I=2\CAS^{SO(5)}-\CAS^{SO(6)} 
\end{equation}
has eigenvalues 
\begin{equation}
\mathrm{C}_I\V^{(n,m)}=2m(m+1)\V^{(n,m)}
\end{equation}
and is precisely the operator we require to separate the
wanted $S^4$ modes from the unwanted modes.
We are therefore in a position to fix the geometry 
for our fuzzy space in a fashion which will suppress the 
non-$S^4$ modes. The desired Laplacian will be
\begin{equation}
\Delta_h=\frac{(\CAS^{SO(6)}+h \mathrm{C}_I)}{2R^2}
\label{delta_h_cas_I}
\end{equation}
The eigenmatrices of this operator are then the polarization 
tensors $\V^{(n,m)}$ and we have the eigenvalue equations
\begin{equation}
\Delta_h \V^{(n,m)}=\left\{\frac{n(n+3) +h m(m+1)}{R^2}\right\}\V^{(n,m)} 
    \label{D104}
\end{equation}
From the spectrum we see that $\Delta_h$ has positive spectrum for 
$h\in (-1,\infty)$ for all values of $L$. In fact for $L$ finite 
the permitted values of $h$ are slightly larger and one
can choose $h\in [-(L+2)/(L+1),\infty)$.

Furthermore now that we have identified a Laplacian type operator which
distinguishes between the $S^4$ and non-$S^4$ modes we can further
identify the projector that removes the unwanted modes from an
arbitrary matrix. It is simply
\begin{equation}
{\cal P}_{S^4}=\prod_{n=1}^{L}\prod_{m=1}^{n}
\frac{\CAS^{SO(5)}-\lambda_{n,m}}{\lambda_{n,0}-\lambda_{n,m}}
=\prod_{m=1}^{L}\frac{2m(m+1)-\mathrm{C}_I}{2m(m+1)}
\label{proj_s4}
\end{equation}
where $\lambda_{n,m}$ are the eigenvalues of $\CAS^{SO(5)}$ of
(\ref{D101}).

As mentioned the choice of $\CAS^{SO(6)}$ as Laplacian fixes the
geometry to be that of a round $\CP^3$. Choosing the linear
combination (\ref{delta_h_cas_I}) of the two Casimirs (\ref{so6_cas})
and (\ref{so5_cas}) still corresponds to a fuzzy approximation of
$\CP^3$, We are doing a continuous deformation of its geometry.  The
set of eigenmatrices and the function algebra is unchanged, only the
Laplacian and its eigenvalues are deformed in a continuous fashion. It
will no longer correspond to a round $\CP^3$ but rather to a squashed
$\CP^3$ with $SO(5)$ rather than $SO(6)$ symmetry.  We will make the
geometry more explicit in a subsequent article
\cite{Brian_Julieta_Denjoe}.

We now see that choosing the modification of the action to be of the
form
\begin{equation}
S_I[\Phi]=\frac{Tr}{d_L}(\Phi \mathrm{C}_I \Phi)
\end{equation}
has precisely the desired properties, i.e. of being zero on matrices
corresponding to functions on $S^4$ and otherwise positive. So when
the parameter $h$ is made arbitrarily large the non-$S^4$ modes are
suppressed.

\vfill\eject
\section{Scalar field theory on fuzzy $4$-sphere}

In this section we will summarize the prescription for working with a
scalar field on $S^4_F$. We begin with the round scalar field theory
on $\CP^3$ given by the action,
\begin{eqnarray}
S_0[\Phi] &=&\frac{R^4}{d_L} Tr \left( \frac{1}{4 R^2} 
[J_{AB},\Phi]^\dag [J_{AB},\Phi] + V[ \Phi ] \right)  \quad .
\label{S_0.1}
\end{eqnarray}
To this we add, the $SO(6)$ non-invariant but $SO(5)$ invariant term
\begin{eqnarray}
  S_I [ \Phi ] &=&\frac{R^4}{d_L} Tr \frac{1}{2R^2}\left( [J_{ab},\Phi]^\dag
[J_{ab} \Phi] - \frac{1}{2} [J_{AB}\Phi]^\dag[J_{AB}\Phi]\right)    \label{S_I.1}
\end{eqnarray}
so that we have the overall action $S[\Phi]=S_0[\Phi]+h S_I[\Phi]$.
This prescription is equivalent to taking the Laplacian which
specifies the geometry to be
\begin{eqnarray}
\Delta_h \cdot & = & \frac{1}{2 R^2}\left(\frac{1}{2}[J_{AB},[J_{AB},
\cdot ]] + h ( [J_{ab}[J_{ab}, \cdot]]-\frac{1}{2} [J_{AB},[J_{AB},
\cdot]])\right) \nonumber\\ &=&
\frac{1}{2R^2}\left([J_a,[J_a,\cdot]]+\frac{(1+h)}{2} ([J_{ab}[J_{ab},
\cdot]]- [J_{a},[J_{a}, \cdot]])\right)
\label{Delta_h}
\end{eqnarray}
or equivalently 
\begin{eqnarray}
 \Delta_h   & = & \frac{1}{2 R^2}\left( \CAS^{SO(6)}  
+ h ( 2\CAS^{SO(5)}-\CAS^{SO(6)}) \right)
\label{Delta_h_cas}
\end{eqnarray}
which gives a stable theory for all $L$ if $h \in (-1,\infty)$. 

This form (\ref{Delta_h_cas}) is an interpolation between $SO(5)$ and
$SO(6)$ Casimirs and the Laplacian is proportional to the $SO(6)$
Casimir for $h=0$ and the $SO(5)$ Casimir for $h=1$. The values of $h$
of interest to us are those large and positive since in the
quantization of the theory following Euclidean functional integral
methods, the states unrelated to $S^4$ then become highly
improbable; this is a direct consequence of (\ref{D104}) and the
expression (\ref{prob_of_config}) for the probability $\mathcal{P}[{\Phi}]$.

Note: we have not specified the potential of the model since the 
above prescription is independent of the potential. The most obvious 
model to consider would be a quartic potential, since this is relevant
to the Higgs sector of the standard model.

\section{Conclusions}

We have presented a solution to the elimination of non-$S^4$ modes in
the fuzzy approach to $S^4$. The solution was to modify the Laplacian
of the scalar action. The modification was to choose the overall
Laplacian to be a perturbation of the round one on $\CP^3$ retaining
only the $SO(5)$ symmetry of $S^4$ and proportional to $\mathrm{C}_I=
2\CAS^{SO(5)}-\CAS^{SO(6)}$.  The resulting Laplacian operator,
which specifies the geometry in the fuzzy models, takes the form
(\ref{Delta_h}) or (\ref{Delta_h_cas}) and has spectrum
\begin{equation}
\lambda_{n,m}=\frac{n(n+3)+hm(m+1)}{R^2}\quad\quad n=0,1,\dots,L \quad \hbox{and}\quad m\leq n\quad .
\end{equation}
It is a positive operator for all $L$ if $h \in (-1,\infty)$.

From our study of the representation content in section 4 it was
possible to construct the projector ${\cal P}_{S^4}$,
eq. (\ref{proj_s4}), which eliminates the non-$S^4$ modes. This is
precisely the projector necessary for the non-associative algebra
discussed in \cite{Ramgoolam_0105006}. We have not pursued this
algebra due to its complications.

As we will see in a subsequent article \cite{Brian_Julieta_Denjoe},
using coherent state techniques it is possible to extract the metric
(and hence the geometry) of the above spaces from the Laplacian. We
will in fact show that the space in all cases is $\CP^3$ and that this
space can be viewed as either the orbit $SU(4)/U(3)$ or the orbit
$Spin(5)/SU(2)\times U(1)$. It therefore in general permits an $SO(5)$
invariant metric when viewed as a $Spin(5)$ orbit, but this can be
expanded to a maximal $SO(6)$ invariance, which we refer to as the
round $\CP^3$. When we have only $SO(5)$ symmetry we have what we call
a squashed $\CP^3$. In fact it is possible to show that $\CP^3$ is
locally the direct product $S^4\times S^2$ and globally an $S^2$
bundle over $S^4$ and that the parameter $h$ is a measure of the size
of the $S^2$ fibres. In fact
\begin{equation}
\frac{R^2_{S^2}}{R^2} = \frac{1}{1+h}
\end{equation}
so the value $h\rightarrow\infty$ corresponds to shrinking the $S^2$ 
fibres to zero size, while $h\rightarrow -1$ corresponds to 
making the fibres infinitely large.

{\bf Acknowledgments:} It is a pleasure to thank A.P. Balachandran,
Brian Dolan, Oliver Jahn, Xavier Martin, Peter Presnajder and Badis
Ydri for helpful discussions.


\end{document}